\documentclass[%
 reprint,
 superscriptaddress,
 amsmath,amssymb,
 aps,
prc,
]{revtex4-2}

\usepackage{color} % Color
\usepackage{graphicx}% Include figure files
\usepackage{dcolumn}% Align table columns on decimal point
\usepackage{bm}% bold math
\newcommand{\Ve}[1]{\ensuremath{\boldsymbol{#1}}}
\newcommand{\be}{\begin{equation}}
\newcommand{\ee}{\end{equation}}
\newcommand{\eq}[1]{\begin{align}#1\end{align}}
\usepackage{braket}
\usepackage[normalem]{ulem}
\begin{document}

%\preprint{APS/123-QED}

\title{Rotational Feshbach resonances in the deformed halo nucleus $^{31}$Ne}% Force line breaks with \\

\author{Shin Watanabe}
\email[]{s-watanabe@gifu-nct.ac.jp}
\affiliation{National Institute of Technology (KOSEN), Gifu College, Motosu 501-0495, Japan}
\affiliation{RIKEN Nishina Center, Wako 351-0198, Japan}

\author{Shoya Ogawa}
\affiliation{Department of Physics, Kyushu University, Fukuoka 819-0385, Japan}

\author{Takuma Matsumoto}
\affiliation{Department of Physics, Kyushu University, Fukuoka 819-0385, Japan}
\affiliation{Research Center for Nuclear Physics (RCNP), Osaka University, Ibaraki 567-0047, Japan}

\author{Kazuyuki Ogata}
\affiliation{Department of Physics, Kyushu University, Fukuoka 819-0385, Japan}

\date{\today}% It is always \today, today,

\begin{abstract}
\noindent{\bf Background:}
The deformed halo nucleus $^{31}$Ne exhibits unique structural properties
 arising from the interplay between its halo neutron
 and the deformation of the $^{30}$Ne core.
While previous studies have established the nature of its weakly-bound ground state
through reaction cross sections and inclusive breakup measurements,
 the structure of its excited states in the low-energy continuum,
 which are expected to appear as resonances, remains largely unexplored.

\noindent{\bf Purpose:}
We investigate the structure of these resonant states
and clarify their rotational nature and formation mechanism.

\noindent{\bf Method:}
The structure of $^{31}$Ne is described using the particle rotor model (PRM),
which explicitly treats the coupling between core excitation and
 single-particle motion in both bound and continuum states.

\noindent{\bf Results:}
The calculations predict the emergence of unbound rotational states
 in the low-energy continuum of $^{31}$Ne, which form a rotational
 sequence built on the weakly-bound Nilsson \mbox{[321\,3/2]} configuration.
 As the total angular momentum increases along the rotational band,
 the intrinsic Nilsson structure is largely preserved,
 whereas the dominant core-spin component shifts to higher spins.
 These resonances are stabilized through the combined effects of
 coupling to higher-lying closed core-excited channels
 and reduced effective neutron relative energies,
 and can therefore be interpreted as rotational Feshbach resonances.

\noindent{\bf Conclusion:}
The present study elucidates the formation mechanism
 of rotational Feshbach resonances in $^{31}$Ne.
This mechanism is expected to be a general feature of weakly-bound deformed nuclei.
Exclusive breakup measurements with coincident detection of $\gamma$ rays
from the de-exciting core will provide crucial tests of
the proposed formation mechanism of rotational Feshbach resonances.

%%%An article usually includes an abstract, a concise summary of the work
%%%covered at length in the main body of the article. 
%%%\begin{description}
%%%\item[Usage]
%%%Secondary publications and information retrieval purposes.
%%%\item[Structure]
%%%You may use the \texttt{description} environment to structure your abstract;
%%%use the optional argument of the \verb+\item+ command to give the category of each item. 
%%%\end{description}
\end{abstract}

\maketitle

%\tableofcontents

%%%%%%%%%%%%%%%%%%%%
%%  Introduction  %%
%%%%%%%%%%%%%%%%%%%%
\section{Introduction}
Nuclei near the drip lines exhibit a wide variety of exotic structures arising
from weak binding and strong coupling to the continuum.
One of the most prominent examples is the formation of neutron halos, in which a
valence neutron extends far beyond the nuclear core~\cite{Tan85,Tan88,Oza01}.
Although halo phenomena were originally studied primarily within
frameworks for spherical nuclei,
it has now been experimentally established that halo structures
can also develop in systems with significant quadrupole deformation of
 the core~\cite{Nak09,Nak14,Kob14,Tak10,Tak12,Tak14}.
Such nuclei are now referred to as \emph{deformed halo} nuclei and provide
 a unique opportunity to explore
 the interplay between single-particle motion and core deformation.

Among the known candidates for deformed halo nuclei,
$^{31}$Ne has attracted particular attention as a key system.
It belongs to the Island of Inversion~\cite{War90}, where $2\hbar \omega$ intruder
 configurations dominate the low-lying states
 and conventional shell closures are quenched.
The last neutron is weakly bound with a very small neutron separation
energy, $S_n=0.15^{+0.16}_{-0.10}$~MeV~\cite{Nak14},
and its core nucleus $^{30}$Ne has a large quadrupole deformation parameter
$\beta_2$ of $0.53\pm0.03$~\cite{Doo16}.
These features provide favorable conditions for the formation of a deformed halo through the
coupling between the valence neutron and the deformed core.

Inclusive reaction measurements, such as
 one-neutron removal cross sections~\cite{Nak09,Nak14}
 and interaction cross sections~\cite{Tak10,Tak12},
 have firmly established the halo nature of $^{31}$Ne.
These studies have significantly constrained its binding properties and
provided valuable information on the dominant configurations of the ground
 state~\cite{Ham10,Min11,Min12,Sum12,Ura11,Ura12,Ura17,Hor10,Hor12,Shu14,Ram15,Tak23}.
In contrast, the structure of excited states of $^{31}$Ne, which are expected to be
predominantly unbound, remains largely unexplored.
Recently, excited states in $^{31}$Ne were observed for the first time
through a two-proton removal reaction from $^{33}$Mg~\cite{Chr21}.
While this experiment provided the first evidence for neutron-unbound excited structures,
 the absence of $\gamma$-ray detection from the de-exciting core prevented the identification
 of the decay channels, such as $^{30}$Ne$(0^+)+n$ and $^{30}$Ne$(2^+)+n$,
 and consequently the spin-parities of the individual resonant states.

In deformed nuclei, rotational excitations naturally arise as a consequence of
 the broken spherical symmetry~\cite{Bohr75}.
 However, in weakly-bound deformed systems such as $^{31}$Ne,
the corresponding rotational excitation energies can be comparable to or larger
than the neutron separation energy, $S_n\sim0.15$~MeV.
As a result, these excited states are expected to strongly couple to
 the continuum and may not form well-defined structures.
This expectation raises the question of whether rotational states
 can persist in such weakly-bound systems.
As will be shown in this work, such states can emerge
 as well-defined resonances in the low-energy continuum,
reflecting the strong coupling between the valence neutron and the core rotation.

Similar questions have been addressed in previous studies, for example in the case
 of the one-neutron deformed halo nucleus $^{11}$Be, where rotational-like structures
 were shown to persist even in the continuum~\cite{Fos16}.
 In this system, the relatively high excitation energy of the core,
 $E_{2^+}(^{10}\mathrm{Be})=3.368$~MeV, leads to significant
 configuration mixing, often associated with strong $K$ mixing~\cite{Fos16},
 where $K$ denotes the projection of the total angular momentum onto the
 symmetry axis. As a result, the rotational structure cannot be understood
 within the simple strong-coupling limit of the particle-rotor picture~\cite{Bohr75},
 in which the core rotates sufficiently slowly so that the valence neutron
 adiabatically follows the rotating deformed potential while preserving
 the intrinsic Nilsson structure.
 In contrast, heavier deformed halo systems such as $^{31}$Ne provide a
 qualitatively different regime. The low excitation energy of the core,
 $E_{2^+}(^{30}\mathrm{Ne})=0.801$~MeV, favors the strong-coupling limit,
 in which the valence neutron is strongly coupled to the rotating deformed core.
In this sense, $^{31}$Ne is a favorable system
for investigating rotational motion in the continuum.
In the present work, we investigate how the intrinsic Nilsson structure can
 persist in the continuum and how it gives rise to rotational resonances.

Recently, advances in experimental techniques at 
the Radioactive Isotope Beam Factory (RIBF) have made exclusive breakup
 measurements of $^{31}$Ne feasible~\cite{Tom17}.
These measurements enable the identification of specific decay channels,
 such as $^{30}$Ne$(2^+)+n$, through the coincident detection of the
 emitted neutron, the core, and the $\gamma$ rays from the de-exciting core.
These observables provide direct access to excited states in the continuum,
and highlight the need for a deeper understanding of the underlying mechanism
 that stabilizes resonant states in weakly-bound deformed nuclei.
From a theoretical perspective, key ingredients in this problem are
 the proper treatment of core excitation and the continuum behavior of the valence neutron.
 The particle rotor model (PRM) provides a natural framework
 for such weakly-bound deformed core-plus-neutron systems by
 treating them in a unified manner~\cite{Bohr75,Ura11,Ura12,Ura17,Wat24_def-PF}.
 In addition, the PRM framework can be naturally incorporated into reaction models that explicitly
 treat core excitation~\cite{Mor12,Mor12-2,Sum06,Sum06-2,Die14,Lay16,Die17,Hag22},
 providing a bridge between nuclear structure and reaction observables.

In this work, we investigate the structure of the unbound rotational states of $^{31}$Ne
 that appear in the low-energy continuum and clarify their formation mechanism.
 Using the PRM, we obtain the structure of these resonant states. Their formation
 mechanism is then discussed from two complementary viewpoints.
From the channel-coupling viewpoint, we investigate how coupling to core-excited
 channels stabilizes the resonances.
From the intrinsic (strong-coupling) viewpoint, we demonstrate that the underlying
 Nilsson intrinsic structure is largely preserved and provide a simple rotational
 picture based on the rotation of the Nilsson state.

This paper is organized as follows. Section~\ref{sec:Theoretical Framework} outlines
 the theoretical framework. Section~\ref{sec:results} presents the numerical results and
 discusses the formation mechanism of rotational Feshbach resonances.
 Finally, Sec.~\ref{sec:summary} summarizes the main conclusions.

%%%%%%%%%%%%%%%%%%%
%%  Formulation  %%
%%%%%%%%%%%%%%%%%%%
\section{Theoretical Framework} \label{sec:Theoretical Framework}
The structure of $^{31}$Ne is described using the PRM,
where the nucleus is treated as a $^{30}\mathrm{Ne}+n$ two-body model with core excitation.
This framework allows us to analyze the interplay between single-particle motion,
collective rotation, and continuum coupling, which are essential for understanding
the emergence of rotational resonances in deformed halo systems.

In the following, we first introduce the PRM Hamiltonian in Sec.~\ref{sec:PRM}.
To gain physical insight into the structure of the system, 
we employ two equivalent formulations.
In the laboratory frame, the wave function is expanded 
in a channel-coupling representation, as described in Sec.~\ref{sec:channel_coupling}.
On the other hand, in the body-fixed frame, the same system is
 described in terms of the strong-coupling representation
based on Nilsson configurations, as discussed in Sec.~\ref{sec:strong_coupling}.
These two representations provide complementary perspectives for
 interpreting  rotational dynamics and the intrinsic structure of the system.
Finally, model setting is summarized in Sec.~\ref{sec:Model}.

\subsection{PRM Hamiltonian}\label{sec:PRM}
The PRM describes $^{31}$Ne
 as a valence neutron coupled to a deformed $^{30}$Ne core.
The total wave function $\Psi_{JM}$ satisfies the Schr\"odinger equation
\be
(H-\varepsilon)\Psi_{JM}=0,\label{eq:sch_prm}
\ee
where $\varepsilon$ is the total energy of the $^{30}$Ne + $n$ system,
and $J$ and $M$ denote the total angular momentum and its projection
onto the laboratory $z$ axis, respectively.
The PRM Hamiltonian is given by
\eq{
H= T_r
+ V_{n\mathrm{c}}({\Ve r},{\Ve \xi})
+ h_{\mathrm{c}}({\Ve \xi}),
\label{eq:Hp_prm}
}
where ${\Ve r}$ denotes the relative coordinate between the valence neutron
and the core, and ${\Ve \xi}$ specifies the orientation of the symmetry axis
of the deformed core.
The operator $T_r$ represents the kinetic energy associated with ${\Ve r}$,
$V_{n\mathrm{c}}$ is the deformed potential between the neutron and the core,
and $h_{\mathrm{c}}$ is the internal Hamiltonian of the core.
The rotational motion of the core is governed by the Schr\"{o}dinger equation
\be
(h_\mathrm{c}-\epsilon_I)\phi_{IM_I}=0,
\label{eq:hc-eig}
\ee
with the rotational energy
\be
\epsilon_I=\frac{\hbar^2}{2\mathcal{J}_\mathrm{c}}I(I+1),\label{eq:epsI}
\ee
where $\phi_{IM_I}$ is the rotational wave function with the core spin $I$ and
its $z$ component $M_I$,
and $\mathcal{J}_\mathrm{c}$ is the moment of inertia of the core.

To eliminate Pauli-forbidden (PF) states corresponding to occupied orbits
in the core, we introduce a pseudopotential
\be
V_{\mathrm{PF}} = \sum_{i=1}^{N_{\mathrm{PF}}} \lambda_i
\ket{\bar{\Psi}^{(i)}} \bra{\bar{\Psi}^{(i)}},
\label{eq:VPF_prm}
\ee
where $\bar{\Psi}^{(i)}$ represents the $i$th PF state,
$N_\mathrm{PF}$ is the total number of PF states,
and $\lambda_i$ is taken to be sufficiently large ($\lambda_i = 10^6$~MeV)
to ensure numerical convergence.
For a deformed nucleus such as $^{31}$Ne, the PF states $\bar{\Psi}^{(i)}$ should 
reflect the intrinsic deformation of the core.
To construct these deformed PF states, we first solve the Schr\"{o}dinger equation
\be
\left(H^{\mathrm{(Nil)}}-\varepsilon_i^{(\mathrm{Nil})}\right)\bar{\Psi}^{(i)}=0,
\ee
where the Nilsson Hamiltonian is defined by omitting $h_\mathrm{c}$ from Eq.~\eqref{eq:Hp_prm}:
\be
H^{\mathrm{(Nil)}}=T_r+V_{n\mathrm{c}}({\Ve r},{\Ve \xi}).
\ee
The eigenstates of $H^{\mathrm{(Nil)}}$ define the Nilsson single-particle states,
 which provide the basis for the strong-coupling
description discussed in Sec.~\ref{sec:strong_coupling}.
The pseudopotential method using deformed PF states
was first introduced in Ref.~\cite{Wat24_def-PF} and was further extended to
incorporate pairing correlations within the Bardeen--Cooper--Schrieffer (BCS)
 formalism~\cite{Pun25}.

\subsection{Channel-coupling representation in the laboratory frame}\label{sec:channel_coupling}
In the laboratory frame, the $^{31}$Ne wave function is expanded as
\be
\Psi_{JM}({\Ve r},{\Ve \xi})
= \sum_c \frac{u_c(r)}{r}\,
\Phi_{c,JM}(\hat{\Ve r},{\Ve \xi})
\label{eq:tot-wf_channel}
\ee
using the channel basis
\be
\Phi_{c,JM}(\hat{\Ve r},{\Ve \xi})
= [\mathcal{Y}_{\ell j}(\hat{\Ve r}) \otimes \phi_I({\Ve \xi})]_{JM},
\label{eq:channel_base_channel}
\ee
where the channel index $c=\{\ell j I\}$ is specified by the orbital angular momentum $\ell$,
the total single-particle angular momentum $j$, and the core spin $I$.
Here, $\mathcal{Y}_{\ell j}$ is the spin-angular function of the valence neutron,
and $j$ is obtained by coupling $\ell$ with the neutron spin $s=1/2$.
The radial wave functions $u_c(r)$ describe the relative motion between the
valence neutron and the core in each channel. They are obtained by solving
the non-local coupled-channel equations including the pseudopotential $V_{\mathrm{PF}}$
as formulated in Ref.~\cite{Wat24_def-PF}, under appropriate boundary conditions.

To identify resonant states, we employ the complex scaling method (CSM)~\cite{Agu71,Aoy06,Myo26}.
In the CSM, the relative coordinate ${\Ve r}$ and its conjugate momentum {\Ve k} are transformed as 
\be
{\Ve r}\rightarrow {\Ve r}e^{i\theta},\qquad {\Ve k}\rightarrow {\Ve k}e^{-i\theta},
\ee
where $\theta$ is the complex scaling angle.
In the present work, resonant states and their complex energies are identified
from the isolated eigenvalues of the complex-scaled Hamiltonian.

\subsection{Strong-coupling representation in the body-fixed frame}\label{sec:strong_coupling}

We next present the strong-coupling representation of the
PRM~\cite{Bohr75}, which is equivalent to the channel-coupling
representation introduced in Eqs.~\eqref{eq:tot-wf_channel}
and~\eqref{eq:channel_base_channel}.
This representation provides a more transparent interpretation
 of both bound and resonant states in terms of intrinsic
 configurations defined in the body-fixed frame.

In the strong-coupling representation, the total wave function is
expanded as
\be
\tilde{\Psi}_{JM}({\Ve r}',\omega)
=\sum_{\ell j}\sum_{\Omega>0}\frac{v_{\nu}(r')}{r'}\tilde{\Phi}_{\nu,JM}
(\hat{{\Ve r}}',\omega),\label{eq:sc_expansion}
\ee
where ${\Ve r}'$ denotes the coordinate of the
valence neutron in the body-fixed frame, and $\omega$ denotes the
Euler angles specifying the orientation of the intrinsic frame
associated with the core.
The index $\nu=\{\ell j\Omega\}$ represents the quantum numbers of the
single-particle state, where $\Omega$ is the projection of the
single-particle total angular momentum onto the symmetry axis.
For the present axially symmetric rotor, $\Omega$ also corresponds to
the projection $K$ of the total angular momentum $J$ onto the symmetry
axis, i.e., $K=\Omega$.
Imposing the $\mathcal{R}$-invariance and adopting the phase convention of
Ref.~\cite{Bohr75}, the strong-coupling basis is written as
\eq{
\tilde{\Phi}_{\nu,JM}(\hat{{\Ve r}}',\omega)
&=\sqrt{\frac{2J+1}{16\pi^2}}
\Bigl[\mathcal{Y}_{\ell j\Omega}(\hat{{\Ve r}}')D^J_{MK}(\omega)\nonumber\\
&\hspace{5mm}
+(-1)^{J+K}\mathcal{Y}_{\ell j\bar{\Omega}}(\hat{{\Ve r}}')D^J_{M-K}(\omega)\Bigr],
\label{eq:sc_basis}
}
where $\mathcal{Y}_{\ell j\Omega}$ is the spin-angular wave function,
$\mathcal{Y}_{\ell j\bar{\Omega}}$ denotes its time-reversed state,
and $D^J_{MK}(\omega)$ is the Wigner $D$ function.

The core rotational Hamiltonian can be rewritten as
\be
h_\mathrm{c}
=\frac{\hbar^2}{2\mathcal{J}_\mathrm{c}}{\Ve I}^2
=\frac{\hbar^2}{2\mathcal{J}_\mathrm{c}}({\Ve J}-{\Ve j})^2
=\frac{\hbar^2}{2\mathcal{J}_\mathrm{c}}({\Ve J}^2+{\Ve j}^2-2{\Ve J}\cdot{\Ve j}),\label{eq:hc_sc}
\ee
where the last term corresponds to the Coriolis interaction,
\be
h_{\mathrm{Coriolis}}
=-\frac{\hbar^2}{\mathcal{J}_{\mathrm{c}}}
\ensuremath{\boldsymbol{J}}\cdot\ensuremath{\boldsymbol{j}}.
\label{eq:h_coriolis}
\ee
This interaction couples the single-particle motion to the collective rotation
 of the entire system and mixes different $\Omega$ components.

In the strong-coupling limit, where the Coriolis mixing is negligible,
$\Omega$ and $K$ become good quantum numbers with $K=\Omega$.
The corresponding intrinsic state is the Nilsson state
\be
\varphi_{\Omega}({\Ve r}')
=\sum_{\ell j}\frac{v_{\ell j\Omega}(r')}{r'}\mathcal{Y}_{\ell j\Omega}(\hat{{\Ve r}}').
\label{eq:nilsson_intrinsic}
\ee
Using Eq.~\eqref{eq:nilsson_intrinsic}, the total wave function in the
strong-coupling limit can be expressed as
\eq{
\tilde{\Psi}^{(0)}_{K,JM}({\Ve r}',\omega)
&=\sqrt{\frac{2J+1}{16\pi^2}}\Bigl[
\varphi_{\Omega}({\Ve r}')D^J_{MK}(\omega)\nonumber\\
&\hspace{5mm}+(-1)^{J+K}
\varphi_{\bar{\Omega}}({\Ve r}')D^J_{M-K}(\omega)\Bigr],
\label{eq:sc_zeroth}
}
where $\varphi_{\bar{\Omega}}$ denotes the time-reversed state of
$\varphi_{\Omega}$.
This expression provides a simple physical interpretation of the
strong-coupling limit, in which the Nilsson intrinsic state
adiabatically follows the collective rotation of the entire system,
described by the Wigner $D$ function.
Consequently, states with different $J$ built on the same intrinsic
Nilsson state can be interpreted as members of the same rotational band.

In this limit, the $J$ dependence of the energy is given by
\be
\varepsilon^{(0)}(J)
=
\varepsilon'_{\Omega}
+
\frac{\hbar^2}{2\mathcal{J}_{\mathrm{c}}}
J(J+1),
\label{eq:vareps0}
\ee
where $\varepsilon'_{\Omega}$ denotes the effective intrinsic energy
of the Nilsson state, including the contributions independent of $J$.
Equation~\eqref{eq:vareps0} provides a useful reference for examining
rotational-band-like structure of the bound and resonant states
obtained in the present PRM calculations.

\subsection{Model setting for $^{31}$Ne} \label{sec:Model}
In this subsection, we summarize the model setting used in the present
calculations.

The neutron-core potential $V_{n\mathrm{c}}$ is
 taken following Refs.~\cite{Ura11, Ham10}, adapted to the present formulation.
It is given by
\eq{
V_\mathrm{def}({\Ve r},{\Ve \xi})
&=V_\mathrm{WS}(r)
-\beta_2 R_0\frac{dV_\mathrm{WS}(r)}{dr}Y_{20}(\theta_{n{\mathrm{c}}}),\label{eq:Vdef}\\
V_\mathrm{\ell s}(r)
&=-F_{\ell s}r_0^2\frac{1}{r}\frac{dV_\mathrm{WS}(r)}{dr},
}
where $\cos{\theta_{n{\mathrm{c}}}}=\hat{\Ve r}\cdot{\Ve \xi}$.
The radial part $V_\mathrm{WS}(r)$ is taken to be the Woods-Saxon form,
\begin{equation}
V_\mathrm{WS}(r)=-\frac{V_\mathrm{WS}^{(0)}}{1+\exp[(r-R_0)/a]} .
\end{equation}
The radius parameter is defined as $R_0 = r_0 A_\mathrm{c}^{1/3}$, with
$r_0 = 1.27$~fm and $A_\mathrm{c} = 30$ being the mass number of the core.
The diffuseness parameter is set to $a = 0.67$~fm.
The spin-orbit interaction is assumed to be spherical for simplicity,
and its strength is fixed to $F_{\ell s} = 0.44$.
The quadrupole deformation parameter is taken as $\beta_2 = 0.55$,
consistent with Ref.~\cite{Nak14}.
The moment of inertia of the core, $\mathcal{J}_\mathrm{c}$ in Eq.~\eqref{eq:epsI},
 is determined so as to
reproduce the experimental $2^+$ excitation energy of $^{30}$Ne,
$\epsilon_2 = 0.801$~MeV~\cite{Doo09}, which yields
$\hbar^2/(2\mathcal{J}_\mathrm{c}) = 0.1335$~MeV.
Finally, the depth of the Woods-Saxon potential $V_\mathrm{WS}^{(0)}$ is adjusted to
reproduce the experimental one-neutron separation energy
of the $3/2^-$ ground state, $S_n=0.15$ MeV~\cite{Nak14}.
With this condition, the potential depth is determined to be
$V_\mathrm{WS}^{(0)} = 41.084$~MeV.

The model space includes the partial waves with $\ell=1$, 3, and 5.
The contributions from higher partial waves ($\ell\ge7$) were found to
 be negligible.
For each total angular momentum $J$, all core-spin states $I$ satisfying
 the angular-momentum coupling conditions were included.
The inclusion of all allowed core-spin states is essential for preserving
the degeneracy of the Nilsson states in the strong-coupling limit.

%%%%%%%%%%%%%%%
%%  Results  %%
%%%%%%%%%%%%%%%
\section{Results and Discussion}\label{sec:results}

\subsection{Structural properties of ${}^{31}$Ne}\label{sec:structure}
\begin{figure}[htbp]
\includegraphics[width=0.45\textwidth,clip]{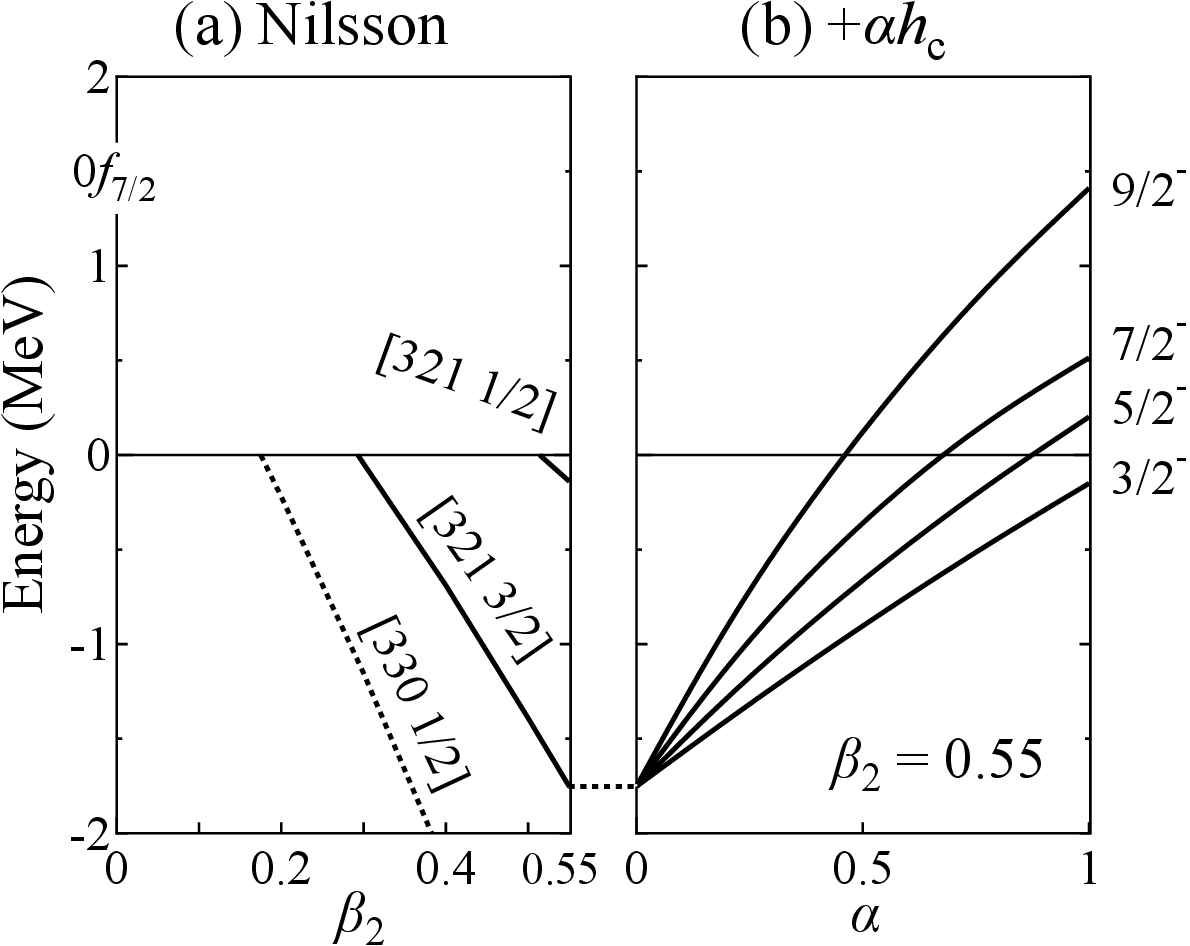}
\caption{The basic mechanism underlying the present work:
 a deformed Nilsson configuration evolves into a rotational-band-like
 sequence once the core-excitation term $h_\mathrm{c}$ is taken into account.
(a) Nilsson diagram as a function of the quadrupole deformation parameter $\beta_2$.
The \mbox{[330\,1/2]} orbital and those below it are Pauli-forbidden, while 
the \mbox{[321\,3/2]} configuration evolves into the physical weakly-bound
or low-lying excited states.
(b) Splitting of the \mbox{[321\,3/2]} manifold by varying the strength
of the core-excitation term $\alpha h_c$  $(0\le\alpha\le1)$,
 leading to the $J^\pi=3/2^-$, $5/2^-$, $7/2^-$, and $9/2^-$ members of
 the $K=3/2$ rotational-band-like sequence.
}
\label{fig:Nilsson_alpha}
\end{figure}

Figure~\ref{fig:Nilsson_alpha} illustrates 
how the low-lying rotational spectrum of $^{31}$Ne emerges from the
Nilsson single-particle configurations.
In the Nilsson diagram [Fig.~\ref{fig:Nilsson_alpha}(a)], all orbitals below the
\mbox{[330\,1/2]} configuration are Pauli forbidden because they correspond to
states already occupied in the deformed $^{30}$Ne core.
Among the available orbitals, the \mbox{[321\,3/2]} level evolves continuously
into the weakly-bound or low-lying excited states.
When the core-excitation term $\alpha h_{\mathrm{c}}$ $(0\le\alpha\le1)$
 is introduced [Fig.~\ref{fig:Nilsson_alpha}(b)], the degeneracy of the
\mbox{[321\,3/2]} manifold is lifted, and the spectrum splits into the
$J^\pi = 3/2^-$, $5/2^-$, $7/2^-$, and $9/2^-$.
For $\alpha = 1$, corresponding to the physical Hamiltonian,
the $3/2^-$ state becomes the ground state with an eigenenergy of
$\varepsilon=-0.15$~MeV, whereas the $5/2^-$, $7/2^-$, and $9/2^-$
 states appear as unbound resonances with $\varepsilon>0$.
These resonant states originate from the Nilsson \mbox{[321\,3/2]}
 configuration with $K=3/2$,
providing the basis for the rotational-band interpretation developed below.

\begin{figure}[htbp]
\includegraphics[width=0.45\textwidth,clip]{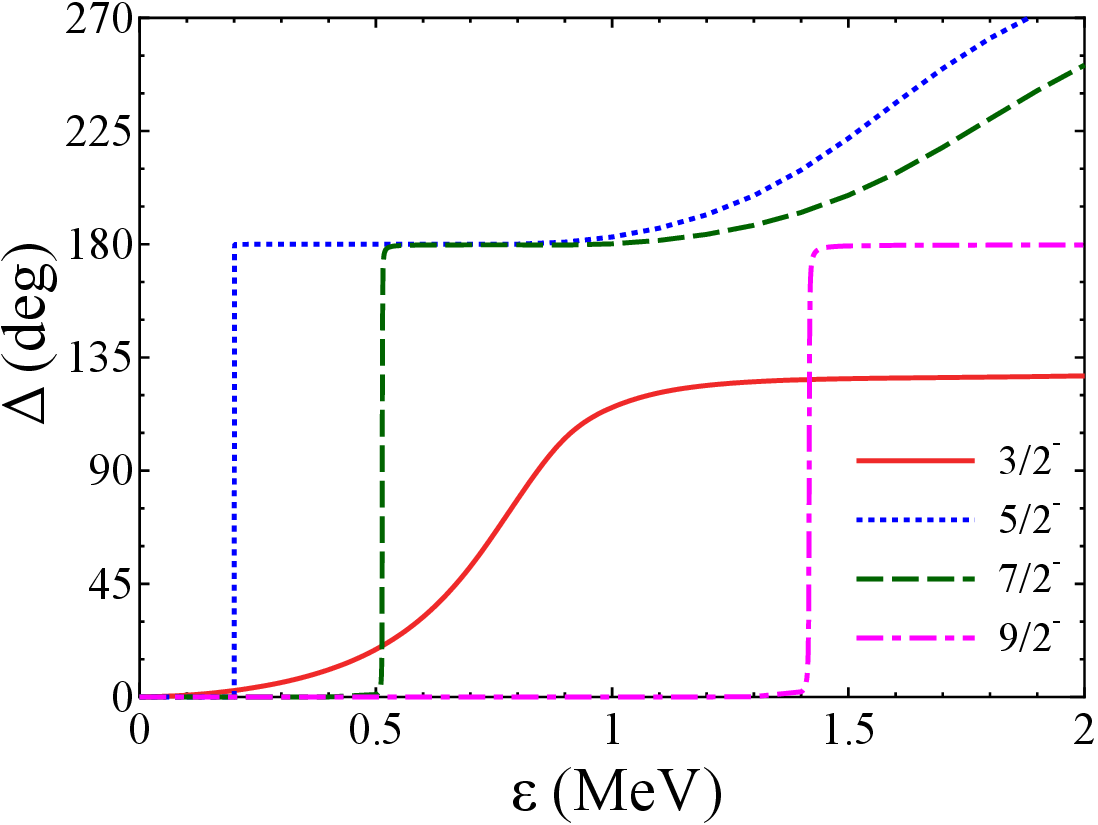}
\caption{Eigenphase sum $\Delta$ as a function of the excitation energy $\varepsilon$.
The solid, dotted, dashed, and dot-dashed curves correspond to
the $J^\pi = 3/2^-$, $5/2^-$, $7/2^-$, and $9/2^-$ states, respectively.
}
\label{fig:eigenps_sum}
\end{figure}

To confirm the resonant character of the low-lying excited states,
we examine the eigenphase sum $\Delta$ obtained by solving the coupled-channel
scattering problem~\cite{Haz79,Hag04}.
As shown in Fig.~\ref{fig:eigenps_sum},
$\Delta$ exhibits an almost step-like behavior, indicating narrow resonances,
 typically with widths of $\Gamma\lesssim0.01$~MeV.
The corresponding resonance energies are consistent with
the rotational spectrum for $\alpha=1$ shown in Fig.~\ref{fig:Nilsson_alpha}(b).

We begin our structural analysis with the ground state of $^{31}$Ne.
As described in Sec.~\ref{sec:Model},
the depth of the neutron-core potential is adjusted to reproduce the
experimental one-neutron separation energy
 of $S_n=0.15$ MeV~\cite{Nak14}.
For comparison, we extracted an experimental RMS matter
 radius of $3.51\pm0.05$~fm for $^{31}$Ne
 from the measured interaction cross section~\cite{Tak12}
 using the method described in Ref.~\cite{Wat14}.
Assuming the experimental RMS matter radius of $3.33\pm0.04$~fm for the $^{30}$Ne core,
the present calculation gives an RMS matter radius of 3.38~fm for $^{31}$Ne.
The reasonable agreement between the calculated and extracted radii supports
 the validity of the present model in describing the
 ground-state structure of $^{31}$Ne.

\begin{figure}[htbp]
\includegraphics[width=0.45\textwidth,clip]{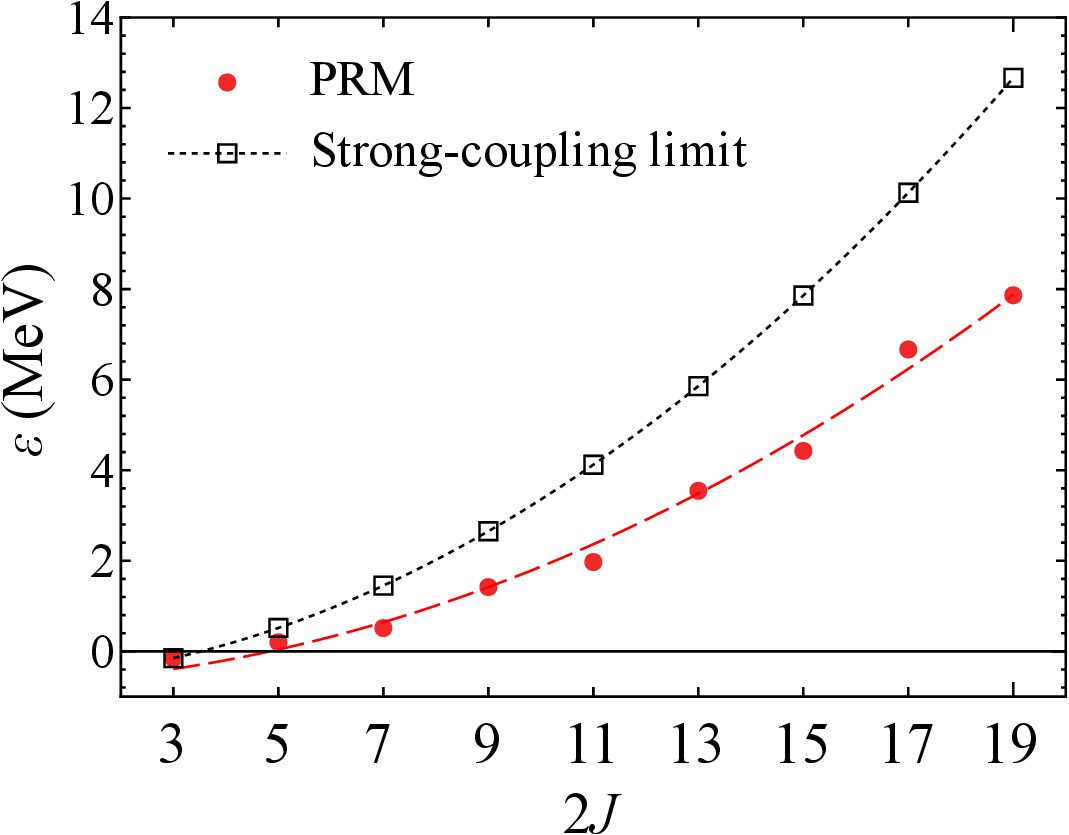}
\caption{Excitation energies of the rotational-band-like states with $K=3/2$
obtained in the full PRM calculation (filled circles), compared with the rigid
strong-coupling limit (open squares with a dotted line).
}
\label{fig:rotE}
\end{figure}

Figure~\ref{fig:rotE} compares the excitation energies of the rotational-band-like
states obtained in the full PRM calculation with those expected from the rigid
strong-coupling limit given in Eq.~\eqref{eq:vareps0}.
While the strong-coupling energies follow the characteristic $J(J+1)$ dependence
with a fixed moment of inertia, the PRM results are systematically lower
for all spin values.
By fitting the PRM excitation energies as a function of $J(J+1)$,
as indicated by the dashed curve, we define an effective moment of inertia
$\mathcal{J}_{\mathrm{eff}}$.
The fit gives $\hbar^2/(2\mathcal{J}_{\mathrm{eff}})=0.0861$~MeV, which corresponds to
$\mathcal{J}_{\mathrm{eff}}/\mathcal{J}_\mathrm{c}\approx1.55$ relative to the rigid
strong-coupling value $\hbar^2/(2\mathcal{J}_\mathrm{c})=0.1335$~MeV.
This result indicates a significant enhancement of the effective moment
of inertia in the PRM compared with the strong-coupling limit.

This enhancement can be associated with dynamical effects such as Coriolis mixing
and coupling to continuum states.
From the viewpoint of the strong-coupling limit,
both the deformed core and the intrinsic Nilsson configuration of the valence neutron
are fixed in the body-fixed frame, while the rotational motion
corresponds to a collective rotation of the entire system.
The Coriolis interaction then mixes Nilsson states with different $\Omega$.
Since the \mbox{[321\,3/2]} state is already weakly bound,
the Coriolis mixing naturally couples it to continuum states.
As the rotational excitation energy increases, the bound Nilsson intrinsic
state evolves continuously into a sequence of unbound rotational resonances
through Coriolis mixing and continuum coupling.
At the same time, these dynamical effects enhance the effective moment of
 inertia and reduce the excitation energies of the rotational states.
The strong-coupling limit therefore provides a useful reference,
while the full PRM calculation incorporates the additional effects
arising from Coriolis mixing and continuum coupling.
Nevertheless, the resulting rotational states remain consistent with
a rotational sequence built on the underlying Nilsson \mbox{[321\,3/2]} state.

In the following analysis, we employ the pseudostates, denoted by $\hat{\Psi}$,
 obtained with the Gaussian expansion method (GEM)~\cite{Hiy03} to investigate the
 intrinsic structures of the resonant states.
In the present work, the resonant states and their resonance energies
 are determined using the CSM, whereas the corresponding pseudostates provide a convenient
 discrete representation of the continuum
 for evaluating overlaps and configuration probabilities.
For each member of the rotational-band-like sequence shown in Fig.~\ref{fig:rotE},
 the corresponding pseudostate can be uniquely identified.
 These pseudostates are well isolated from the surrounding nonresonant
 pseudostates and remain stable against variations of the GEM basis parameters.
 The structural properties extracted from these pseudostates are therefore expected to
 provide reliable representations of the corresponding resonant states.

\begin{figure}[htbp]
\includegraphics[width=0.45\textwidth,clip]{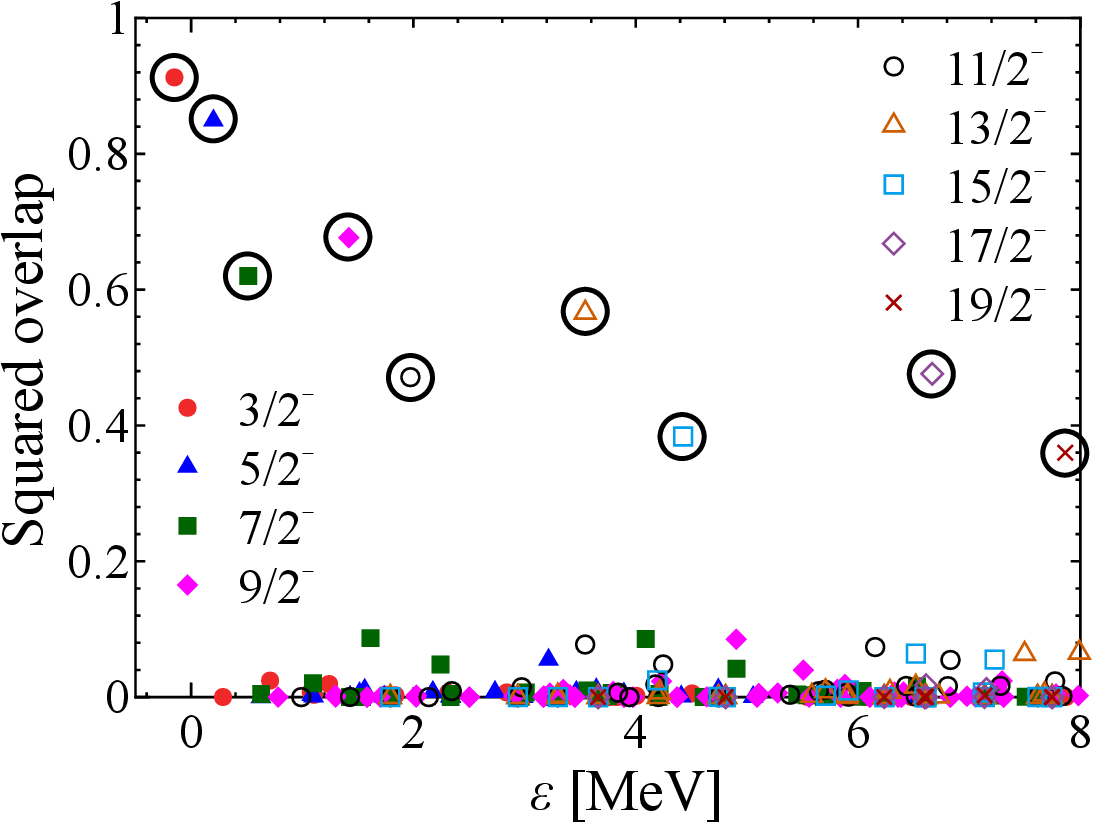}
\caption{Squared overlap between the Nilsson intrinsic state obtained at $\alpha=0$
 and the PRM pseudostates at $\alpha=1$, $|\braket{\Psi_{\rm Nil}|\hat{\Psi}_{J^\pi}}|^2$,
 as a function of the excitation energy $\varepsilon$.
Different symbols correspond to different spin-parity values.
The states identified as members of the rotational-band-like sequence are
highlighted by circles.
}
\label{fig:over_all}
\end{figure}

Figure~\ref{fig:over_all} shows the squared overlap between the Nilsson intrinsic
state obtained at $\alpha=0$ and the PRM pseudostates at $\alpha=1$,
$|\braket{\Psi_{\rm Nil}|\hat{\Psi}_{J^\pi}}|^2$, as a function of the excitation energy $\varepsilon$
with respect to the $^{30}$Ne$(0^+)+n$ threshold energy ($\varepsilon=0$~MeV).
For each $J^\pi$, only one state exhibits remarkably large overlaps
with the Nilsson state \mbox{[321\,3/2]}.
These states form a rotational-band-like sequence built on the same intrinsic structure.
While the $3/2^-$ state appears as a bound state, the higher-spin
members ($5/2^-$, $7/2^-$, $9/2^-$, $\ldots$) emerge as pseudostate
solutions in the continuum.
Although the overlap gradually decreases with increasing spin,
 a single state for each $J^\pi$ continues to exhibit a significantly
larger overlap than those of the other pseudostates,
 indicating that the underlying Nilsson intrinsic structure
 persists even at higher excitation energies.
In contrast, the remaining pseudostates show only negligible overlaps,
typically below $0.1$, demonstrating that they are essentially orthogonal
 to the Nilsson intrinsic state and do not participate
in the rotational-band-like structure.
These selective overlaps therefore support an interpretation based on
the strong-coupling picture, in which the intrinsic quantum number
$K$ is approximately conserved and the rotational states
originate from a common Nilsson intrinsic state.

\begin{figure}[htbp]
\includegraphics[width=0.45\textwidth,clip]{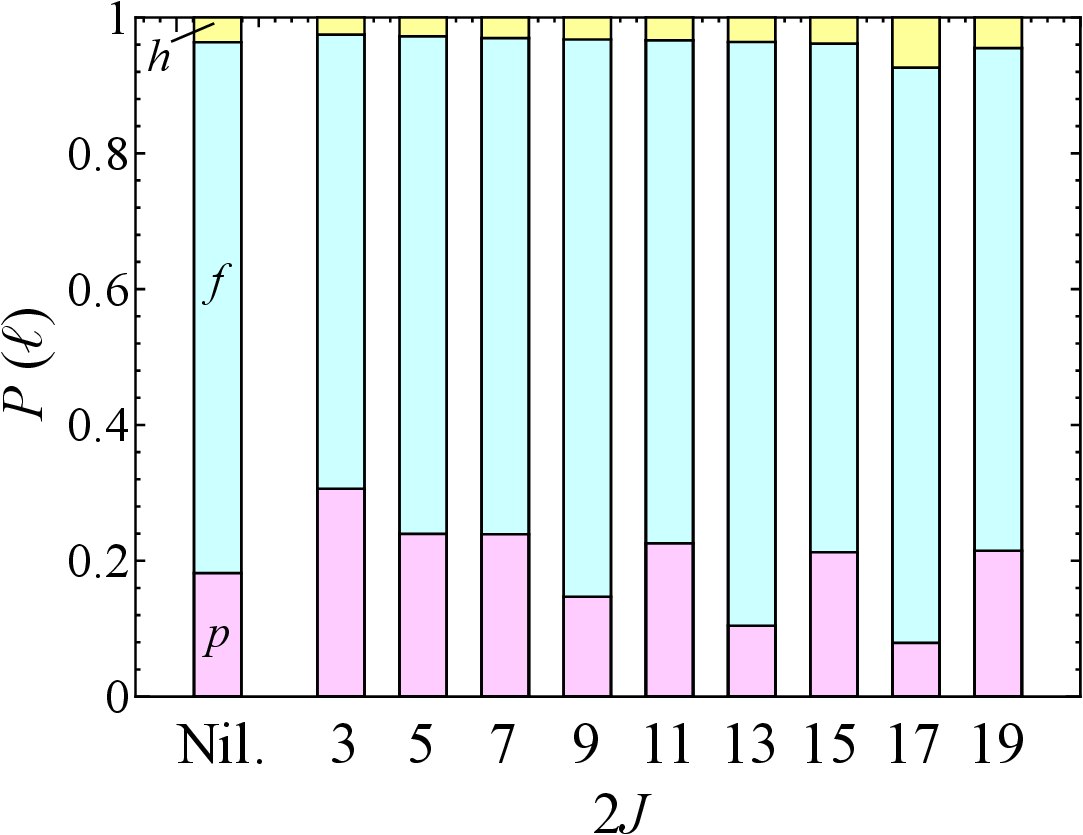}
\caption{
Orbital-angular-momentum probabilities $P(\ell)$ of the valence neutron
for the low-lying rotational-band-like states in $^{31}$Ne.
The contributions from $\ell=1$, $3$, and $5$ are shown separately.
The leftmost bar corresponds to the Nilsson \mbox{[321\,3/2]} state for comparison.
}
\label{fig:prob_ell}
\end{figure}

Figure~\ref{fig:prob_ell} shows the orbital-angular-momentum probabilities $P(\ell)$
 of the valence neutron for the rotational-band-like states.
 The orbital composition remains remarkably close
 to that of the intrinsic Nilsson \mbox{[321\,3/2]} state over the
 entire rotational band. This demonstrates that the underlying single-particle
 structure is largely preserved even in the continuum.
 This result is consistent with the large overlaps shown in Fig.~\ref{fig:over_all}
 and supports the strong-coupling interpretation discussed above.

\begin{figure}[htbp]
\includegraphics[width=0.45\textwidth,clip]{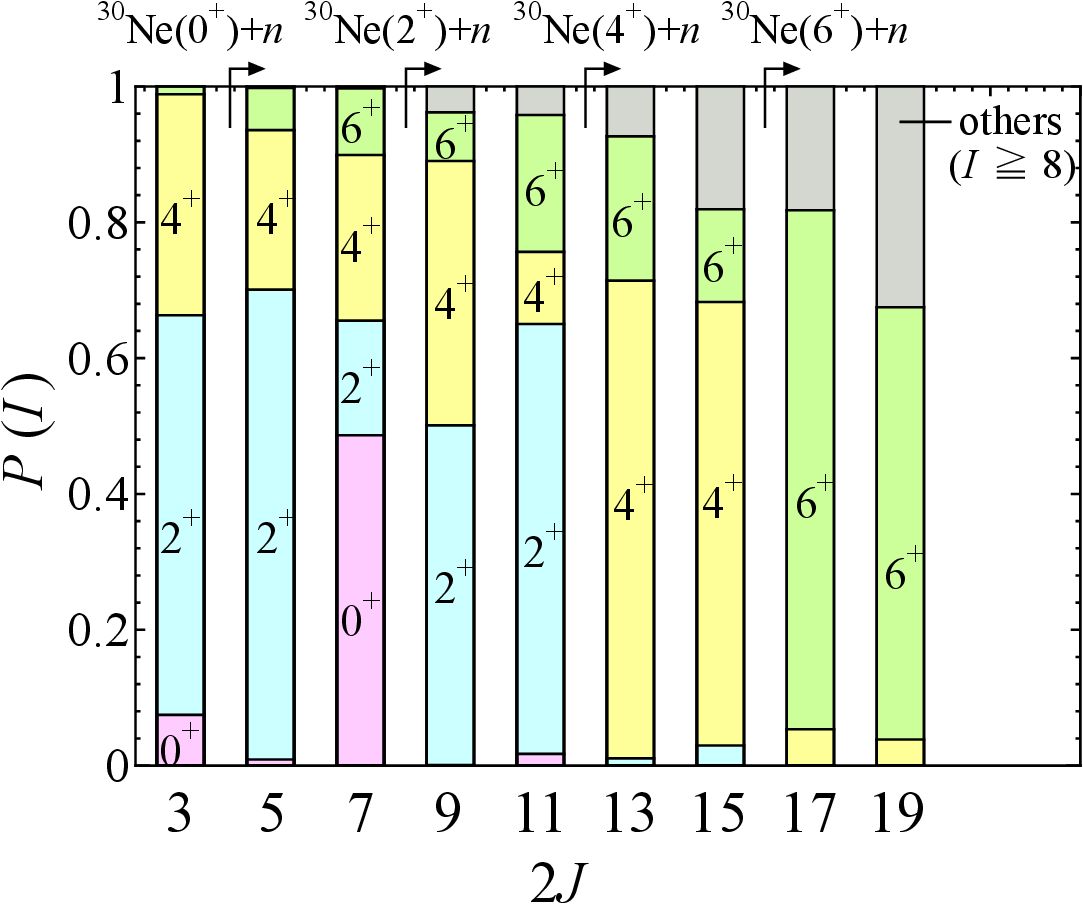}
\caption{
Core-spin probabilities $P(I)$ of the low-lying rotational-band-like states in $^{31}$Ne.
The contributions from $I^\pi=0^+$, $2^+$, $4^+$, $6^+$ are shown separately,
while the small components with $I\ge8$ are summed as ``others'' in the figure.
The openings of the $^{30}$Ne$(I^+)+n$ channels are indicated by the arrows.
}
\label{fig:prob024}
\end{figure}

Figure~\ref{fig:prob024} shows the core-spin probabilities
 $P(I)$ for the rotational-band-like states.
For the ground state ($3/2^-$), the contribution from the core-excited
configurations clearly dominates over the $0^+$ configuration,
indicating that the halo structure is largely associated with
core-excited components.
In general, core excitation increases the effective binding energy
of the valence neutron and thus suppresses halo formation.
However, since the $2^+$ excitation energy of the $^{30}$Ne core is relatively low,
 $\epsilon_{2^+}=0.801$~MeV, the effective binding energy
 of the dominant $2^+$ configuration remains only about $0.95$~MeV.
This indicates that the halo structure can be maintained despite
the strong coupling to the low-lying collective states.
The coexistence of weak binding and strong core excitation may
 be a general feature of heavy deformed halo nuclei such as $^{37}$Mg.

Among the excited states, the $5/2^-$ state is dominated by the $2^+$ core component,
 while the contribution from the $0^+$ core component is negligible.
 This state represents a typical Feshbach resonance,
 in which the resonance is stabilized primarily through coupling to the closed
 $^{30}$Ne$(2^+)+n$ channel. In contrast, the $7/2^-$ state is dominated by
 the $0^+\otimes f_{7/2}$ configuration (49\%), which can form a single-particle
 resonance owing to the centrifugal barrier. Nevertheless, the remaining 51\%
 of the wave function consists of core-excited components, indicating that
 coupling to the closed core-excited channels further enhances the stability
 of the resonance.

As the total angular momentum $J$ increases further, the dominant core-spin component
 evolves systematically from $2^+$ to $4^+$ and then to $6^+$,
 while the underlying Nilsson structure is largely preserved.
Although these dominant configurations correspond to open channels,
the increase in the core-excitation energies reduces the effective relative energies
 of the neutron, thereby favoring resonance stabilization.
 In addition, admixtures of higher-lying closed core-excited components further
 suppress neutron decay. These combined effects stabilize the rotational
 resonances as the rotational band develops. 
 The low-lying resonances in $^{31}$Ne can therefore be interpreted
 as rotational Feshbach resonances in the continuum.

\subsection{Stability of unbound rotational states against potential uncertainty}

We now return to the CSM analysis to examine the robustness of the predicted resonances
 against uncertainties in the neutron-core interaction.
In realistic descriptions of unstable nuclei,
the neutron-core interaction inevitably involves a certain degree of uncertainty.
In particular, for deformed halo systems such as $^{31}\mathrm{Ne}$,
variations of the effective neutron-core potential at the level of
10--20\% are not unrealistic, reflecting ambiguities in the core structure,
density distribution, and effective interaction.
It is therefore important to examine how robust the predicted resonant structures
are against such variations in the model parameters.

\begin{figure}[htbp]
\includegraphics[width=0.45\textwidth,clip]{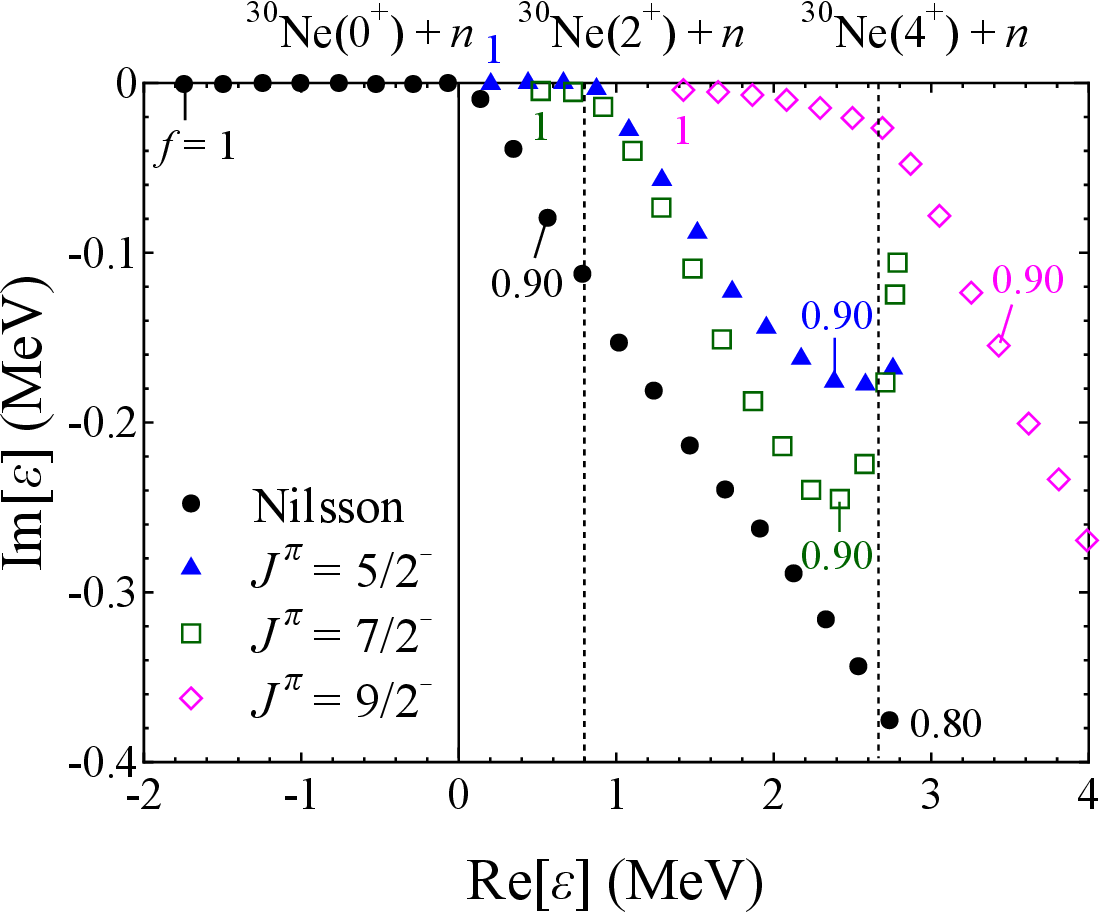}
\caption{
Trajectories of the low-lying states of $^{31}$Ne in the complex-energy plane.
Each trajectory is obtained by gradually reducing
 the depth of the neutron-core potential from
its physical value ($f=1$) in steps of 0.01.
The numbers attached to the trajectories indicate the corresponding values
of the potential scaling factor $f$.
The trajectory of the Nilsson single-particle state [321\,$3/2$] is shown for
comparison.
The vertical lines indicate the $^{30}$Ne$(0^+)+n$ and $^{30}$Ne$(2^+)+n$
 thresholds together with the estimated $^{30}$Ne$(4^+)+n$ threshold
 obtained from Eq.~\eqref{eq:epsI}.
}
\label{fig:res_trace}
\end{figure}

Figure~\ref{fig:res_trace} shows the trajectories of the low-lying states
of $^{31}\mathrm{Ne}$ in the complex-energy plane
obtained by gradually reducing the potential scaling factor $f$
 from its physical value $f=1$ in steps of 0.01.
For comparison, the trajectory of the Nilsson single-particle state
$[321\,3/2]$ is also shown.
Because this Nilsson configuration contains a significant $f_{7/2}$ component,
a single-particle resonance is formed above the $^{30}\mathrm{Ne}(0^+) + n$
threshold owing to the centrifugal barrier.
However, the imaginary part $|\mathrm{Im}[\varepsilon]|$
increases almost linearly with $\mathrm{Re}[\varepsilon]$,
indicating a gradual loss of stability
as the state increasingly couples to the open decay channel.

In contrast, the rotational states with $J^\pi = 5/2^-$, $7/2^-$,
and $9/2^-$ exhibit a markedly different behavior.
Even above the $^{30}\mathrm{Ne}(0^+) + n$ threshold,
their imaginary parts $|\mathrm{Im}[\varepsilon]|$ remain significantly
 smaller than that of the Nilsson state,
demonstrating a significant stabilization against neutron decay.
This enhanced stability originates from the coupling to core-excited channels,
which suppresses the decay into the open $^{30}\mathrm{Ne}(0^+) + n$ channel.
As the excitation energy increases further,
higher core-excited channels such as $^{30}\mathrm{Ne}(4^+) + n$
become involved, particularly for the $9/2^-$ state.
These additional closed channels continue to stabilize the resonances
even above the $^{30}\mathrm{Ne}(2^+) + n$ threshold.
These results clearly demonstrate that the low-lying states of $^{31}\mathrm{Ne}$
are not simple single-particle resonances but are stabilized by the
 collective excitations of the deformed core.
Such a stabilization mechanism is expected to be a general feature
of weakly-bound deformed core-plus-neutron systems,
including heavier deformed halo nuclei such as $^{37}\mathrm{Mg}$.

It is also found that even a 3\% reduction in the neutron-core
 potential depth increases the excitation energies of the $5/2^-$
and $7/2^-$ states sufficiently to open the
$^{30}\mathrm{Ne}(2^+) + n$ channel.
Once this channel becomes energetically accessible,
$|\mathrm{Im}[\varepsilon]|$ increases rapidly,
indicating the rapid onset of decay into the newly opened core-excited channel.
This behavior has important implications for reaction observables.
The opening of the $^{30}\mathrm{Ne}(2^+) + n$ channel means that,
for realistic variations of the neutron-core potential,
the breakup cross sections contain significant contributions
from both the $^{30}\mathrm{Ne}(0^+) + n$
and $^{30}\mathrm{Ne}(2^+) + n$ channels.
Consequently, the channel-resolved breakup cross sections are sensitive
to the proximity of the core-excited thresholds.
Accordingly, measurements without coincident $\gamma$-ray detection
 cannot distinguish between the $^{30}$Ne$(0^+)$ and $^{30}$Ne$(2^+)$ channels.
Coincident detection of $\gamma$ rays from the de-exciting $^{30}\mathrm{Ne}$ core
is therefore essential to disentangle different configurations
and thus provides a direct experimental test of
the proposed formation mechanism of rotational Feshbach resonances.

%%%%%%%%%%%%%%
%  Summary  %%
%%%%%%%%%%%%%%
\section{Summary}\label{sec:summary}

In this work, we have investigated the structure of
 the unbound rotational states in the deformed halo nucleus
 $^{31}\mathrm{Ne}$ using the particle rotor model (PRM).
 The calculations predict a rotational-band-like sequence built
 on the Nilsson $[321\,3/2]$ configuration,
 in which the $3/2^-$ ground state is bound,
 while the higher-spin members appear as narrow resonances in the continuum.
 The large squared overlaps between these states and the intrinsic Nilsson
 configuration indicate that they robustly preserve the underlying Nilsson
 intrinsic structure even above the particle-decay thresholds.

The present study clarifies the formation mechanism of these rotational resonances.
As the total angular momentum $J$ increases along the rotational band,
 the dominant core-spin component shifts successively to higher spins,
 while the intrinsic Nilsson structure is largely preserved.
 As a result, coupling to higher-lying closed core-excited channels provides
 the essential stabilization against neutron decay. In addition, the increase
 in the core-excitation energies effectively lowers the relative energies
 of the valence neutron, which further stabilizes the resonant states.
 These combined effects stabilize the rotational resonances and
 naturally lead to their interpretation as rotational Feshbach resonances.
 This interpretation is further supported by the complex-energy trajectories,
 which show that the rotational resonances remain substantially narrower
 than the corresponding Nilsson single-particle resonance over a wide energy region.

The mechanism identified here is expected to be a general feature of
weakly-bound deformed nuclei and may also be realized
in heavier deformed halo nuclei such as $^{37}$Mg.
Exclusive breakup measurements with coincident detection of $\gamma$
rays from the de-exciting $^{30}$Ne core will provide a direct
experimental test of the proposed formation mechanism of
 rotational Feshbach resonances.

\begin{acknowledgments}
We would like to thank M. Kimura for valuable discussions.
This work was supported
by JSPS KAKENHI Grant Numbers JP26K07076 and JP25K07302,
and by JST ERATO Grant Number JPMJER2304.
\end{acknowledgments}

\bibliography{./ref}% Produces the bibliography via BibTeX.

\end{document}